\documentclass[preprint]{aastex}
\usepackage{epsf}

\newcommand{\om}{\Omega_M}
\newcommand{\ol}{\Omega_\Lambda}

\begin{document}

\slugcomment{\footnotesize{Accepted for publication in {\it The Astronomical Journal} [31 Jan 2001]}}

\title{A Deep Multicolor Survey VII. Extremely Red Objects and Galaxy Formation
\altaffilmark{1}}

\author{Paul Martini\altaffilmark{2}}

\altaffiltext{1}{Based on observations obtained at
MDM Observatory, operated by Columbia University, Dartmouth College, the
University of Michigan, and the Ohio State University} 

\altaffiltext{2}{Current Address: Carnegie Observatories, 813 Santa Barbara St., Pasadena, CA 91101, martini@ociw.edu}

\affil{Department of Astronomy, Ohio State University, 140 W. 18th Ave.,
Columbus, OH 43210} 

\begin{abstract}

Extremely Red Objects (EROs) offer a window to the universe at $z \sim 1$ 
analogous to that provided by the Lyman Break galaxies at $z = 3$. Passive 
evolution and hierarchical galaxy formation models make very distinct 
predictions for the $K$ ($2.2\mu$m) surface density of galaxies at $z \sim 1$ 
and EROs are a powerful constraint on these theories. I present a study of 
nine resolved EROs with $R-K\geq5.3$ and $K\leq18$ mag found in the 
185~arcmin$^2$ 
of the Deep Multicolor Survey with near-infrared imaging. Photometric redshifts 
for these galaxies shows they all lie at $z = 0.8 - 1.3$. The relatively blue 
$J-K$ colors of these galaxies suggest that most are old ellipticals, rather 
than dusty starbursts. The surface density of EROs in this survey ($>0.05$ 
arcmin$^{-2}$), which is a lower limit to the total $z \sim 1$ galaxy surface 
density, is an order of magnitude below the prediction of passive galaxy 
evolution, yet over a factor of two higher than the hierarchical galaxy 
formation prediction for a flat, matter-dominated universe. A flat, 
$\Lambda-$dominated universe may bring the hierarchical 
galaxy formation model into agreement with the observed ERO surface density. 

\end{abstract}


\section{Introduction} \label{sec:intro}

The most well-known, modern method of using color selection to find
high-redshift galaxies is the Lyman-break technique \citep[e.g.,][]{steidel93}
and this method has isolated large numbers of galaxies at $z \sim 3$ and
$z \sim 4$ \citep{steidel96,steidel99}. The great efficiency of the Lyman-break
technique in selecting high-redshift candidates for follow-up spectroscopy
from visible wavelength colors has made it 
easier to preselect samples of $z \sim 3$ galaxy candidates than galaxies at
$z \sim 1 - 2$. A complimentary technique to preselect $z \sim 1$
galaxy candidates is to use a combination of near-infrared (NIR) and visible
wavelength photometry to search for Extremely Red Objects (EROs), which have
large visible--NIR colors \citep{elston88,mccarthy92}.

A sample of galaxies at $z \sim 1$ can test galaxy evolution theories by
measuring both the surface density of these galaxies and the relative
fraction of different spectrophotometric types. Hierarchical galaxy formation
predicts that present day massive galaxies assemble between $z \sim 1$ and
the present \citep[e.g.][]{white78,white91,lacey93,baugh98}. In contrast, the
passive evolution model postulates that galaxies assembled at $z > 3$ and the
comoving space density of bright galaxies at $z \sim 1$ should be comparable
to the local value \citep[e.g.][]{tinsley77,bruzual80}.
\citet{kauffmann98} showed that hierarchical galaxy formation predicts a
significantly smaller fraction of $z \sim 1$ galaxies in a $K$-selected
redshift survey relative to passive galaxy evolution.
$K-$band selection is particularly sensitive to differences between these two
galaxy formation models because it measures the old stellar population that
dominates the mass, rather than recent starbursts which may enhance the
luminosity at visible wavelengths.  Several $K$-selected redshift surveys
have indeed found smaller numbers of $z \geq 1$ galaxies than predicted by
passive evolution, in support of the hierarchical galaxy formation picture
\citep{songaila94,cowie96,fontana99}.

Well-studied EROs to date are all at $z \geq 0.8$ and appear to be either old
ellipticals or dusty starbursts. However, EROs are generally classified 
as ``ellipticals'' because their rest-frame 
visible spectrum is dominated by an old stellar population, rather than on 
the basis of kinematic or surface brightness criteria. 
Old ellipticals at these redshifts have
very red colors because the 4000 \AA\ break falls between the $R$ and $K$
bands. Dusty starbursts have similarly extreme colors due to a combination
of the Balmer continuum break and the relative suppression of the $UV$ light 
from young stars by dust.
The first deep sky survey at $K$ by \citet{elston88} found 2 EROs, which
additional photometry and spectroscopy showed were old ellipticals at
$z = 0.8$ \citep{elston89}. Several more EROs were found by \citet{mccarthy92}
in a $K$ imaging survey of the fields of high-redshift radio galaxies and
two EROs were found by \citet{hu94} in an imaging survey of a $z = 3.790$
quasar field. One of these, HR10, has a spectroscopic redshift of $z = 1.44$
and has since been detected at radio \citep{graham96} and submillimeter
\citep{cimatti98,dey99} wavelengths. These observations provide strong 
evidence that HR10 is a
dusty starburst galaxy. Another well-studied ERO, LBDS 53W091, was discovered
in a NIR study of a sample of weak radio sources. Spectroscopy by
\citet{spinrad97} showed that it is an old, red galaxy at $z = 1.55$.
\citet{soifer99} observed the ERO Cl~0939+4713B, serendipitously discovered
by \citet{persson93}, and found it is an old elliptical at $z = 1.58$.
The spectroscopic observations of HR10, LBDS 53W091, Cl~0939+4713B, and
others show that EROs are at $z \geq 0.8$ and either appear as old
ellipticals or dusty starbursts.

The definition of what constitutes an ERO is highly variable and cuts in color 
include $R-K\geq5$, 5.3, and 6 and $I-K\geq4$. Recent surveys for EROs have
mapped large areas of the sky at NIR wavelengths to statistically study the
ERO population \citep{thompson99,yan00,scodeggio00,daddi00}. \citet{cimatti99}
spectroscopically followed up a sample of nine EROs and found that two are
dusty starbursts and the remaining seven are consistent with passively
evolved ellipticals. \citet{cohen99} have obtained spectra of four of the 19
EROs in the Caltech Faint Galaxy Redshift Survey and all appear to be
ellipticals with little dust. This spectroscopic work indicates that most 
EROs are ellipticals, rather than dusty starbursts. 
While the larger ERO samples only include imaging, \citet{daddi00} showed that 
the EROs in their 700 arcmin$^2$ survey are strongly clustered and their 
surface density can be used to test galaxy formation models. The clustering 
of EROs helps to explain the dispersion in measurements of the ERO surface 
density for a fixed color and magnitude threshold. 

This survey for EROs is based on the visible and NIR imaging of the Deep
Multicolor Survey (DMS) described in \citet{hall96a} and \citet{martini01}. 
The original DMS included $UBVRI_{75}I_{86}$ photometry of 6 high-galactic 
latitude fields over a total area of 0.83 deg$^{2}$. The DMS has been used to 
study the luminosity function of quasars \citep{hall96b,kennefick97}, galactic 
stars \citep{martini98}, the luminosity function of galaxies \citep{liu98}, 
and NIR $JHK$ number counts \citep{martini01}. In this paper I discuss the 
nature of the nine objects with $R-K\geq5.3$ and $K\leq18$ mag in 
the DMS and the implications of the ERO surface density for galaxy evolution 
models. \S\S \ref{sec:phot} and \ref{sec:select} briefly describe the 
photometry and object selection. In \S \ref{sec:analysis} I use colors and 
model fits to the 9-filter spectral energy distributions (SEDs) to compute 
redshifts and classify the EROs as either ellipticals or dusty starbursts. 
These results are discussed in \S \ref{sec:dis}.

\section{Photometry} \label{sec:phot}

The observations, data reduction, and photometric solutions for these
observations are described in \citet{hall96a} and \citet{martini01}. 
To measure the objects detected in the NIR frames (plate scale of $0.3''$ 
pix$^{-1}$) on the CCD data ($0.529''$ pix$^{-1}$), I used the DMS stellar 
catalog \citep{osmer98b} to solve for the coordinate transformations using 
GEOTRANS in IRAF\footnote{IRAF is distributed by the National Optical Astronomy 
Observatories, which are operated by the Association of Universities for 
Research in Astronomy, Inc., under cooperative agreement with the National 
Science Foundation.}.  
The photometric zeropoints for each of the 21 fields was determined 
with the DMS stellar catalog for the CCD images and the photometric 
solutions from \citet{martini01} for the NIR images. Because the subfields with 
NIR data are small, variations in the PSF across the CCD fields described by 
\citet{hall96a} are negligible over these individual subfields. For the 
CCD fields with more than one NIR subfield, the variation in the photometric 
zeropoint for different regions of the same CCD image was $0.02$ mag or 
less.

I measured the brightness of each object using aperture photometry
and a stellar-profile aperture correction. The main purpose of these
measurements is to accurately determine the colors of the galaxies, rather
than their total integrated brightness in each filter.
The same size aperture will sample the same physical region in each galaxy and
provide a better estimate of the shape of the SED than methods such as
isophotal magnitudes, which will depend on the isophotal limit in each filter.
Variations in seeing from filter to filter can complicate this issue as a
larger fraction of the light within a fixed physical radius will fall outside
the aperture if the seeing FWHM is larger. The optimal aperture for these
measurements will be a compromise between a decrease in aperture size, which 
will increase the signal-to-noise ratio, and an increase in the uncertainty 
of the stellar aperture correction. While the image quality is 
on average superior in the NIR frames relative to the CCD data, the NIR 
fields are significantly smaller (9 arcmin$^2$ compared to 225 arcmin$^2$) 
and generally only have one or two stars of sufficient brightness to 
determine the seeing and aperture correction. The NIR frames were therefore 
the limiting factor in minimizing the aperture size. To derive the 
optimal aperture I measured the aperture correction in the NIR frames at a 
range of radii from $0.5''$ to $4''$. I found that for $R<2''$ the 
uncertainty in the aperture correction for many of the frames was 
comparable or greater than the photometric errors at $K\sim18$ mag. I 
therefore chose to use an $R=2''$ aperture for the photometry. 

To measure the amount of lost light from a galaxy in excess of the stellar 
aperture correction, I simulated measurements of exponential disks and 
$r^{1/4}$ profiles with half-light radii $r_h = 0.25''$, $0.5''$, $0.75''$, 
and $1''$ over the range of seeing from $1''$ to $2.5''$ FWHM exhibited by 
these images following the procedure described in \citet{martini01}. This 
range of possible ERO angular sizes span that found by \citet{moriondo00} 
in their study of EROs with {\it HST} imaging. 
For an $R=2''$ aperture, the galaxy light lost in addition to the stellar 
aperture correction ranges from essentially zero for $r_h = 0.25''$ to 
0.2 mag for an $r_h = 1''$ exponential disk and 0.25 mag for an $r_h = 1''$ 
$r^{1/4}$ profile. 

\section{Selection Criteria} \label{sec:select}

NIR imaging surveys for EROs have defined numerous selection
criteria \citep{thompson99,yan00,scodeggio00}, including objects
with $R-K\geq6$, $R-K\geq5$, $I-K\geq4$, and $K$ magnitude upper limits from
18 to 20 mag. An extreme red color is a fairly efficient means of
selecting candidate galaxies at $z \sim 1$ because it brackets the 
4000 \AA\ break in ellipticals and the Balmer continuum plus reddening
in dusty starburst galaxies. This color space could also be inhabited by
late-type stars, emission line galaxies, or very high redshift ($z > 7$)
galaxies, where the Lyman continuum falls between $R$ and $K$. Stellar
contamination is probably the dominant contaminant in ERO samples and can be
avoided by including only resolved sources. The upper limit in magnitude, which
is approximately the spectroscopic limit of the largest current
telescopes, avoids ``contamination'' of the $z \sim 1$ sample by
galaxies at higher redshift as they have much
lower surface densities than $z \sim 1$ objects at these magnitude limits.
For example, a $K = 18$ mag galaxy at $z = 1$ has an absolute magnitude
$M_K = -25$ + log $h$ ($\om = 0.3$, $\ol = 0.7$), two magnitudes brighter 
than $M_{*,K}(z=0) = -23$ + log $h$ \citep{gardner97}. 
The $z \geq 0.8$ galaxies in a survey to
$K \sim 18$ mag will therefore mostly be at $z < 2$ due to the exponential 
decline in the galaxy luminosity function.

For this study I have adopted $R-K \geq 5.3$, $K \leq 18$ mag to select for
EROs in the DMS sample. The $K \leq 18$ mag limit is brighter than at least 
the $r_h = 0.25''$ completeness limit for nearly all 185 arcmin$^2$ of the 
$K$ survey \citep{martini01}.  $R-K \geq 5.3$ was
suggested by \citet{pozzetti00} based on models of ellipticals and dusty
starbursts as well as observations of known EROs at $z \sim 1$. In total,
nine resolved objects in the DMS sample met these two selection criteria for 
a surface density of 0.05 arcmin$^{-2}$. This surface density is a lower 
limit due to the variation in detection efficiency for different galaxy 
sizes and from field to field.  The photometry for the EROs is 
listed in Table~\ref{tbl:phot}.
The only correction that has been applied to the apparent magnitudes in
Table~\ref{tbl:phot} is a stellar aperture correction. 
As these measurements may underestimate the total integrated brightness, this
survey may underestimate the true ERO surface density (see \S~\ref{sec:dis}).

\section{Analysis} \label{sec:analysis}

\subsection{ERO Colors} \label{sec:colors}

\begin{figure*}[t]
\centerline{
\epsfxsize=3.5truein\epsfbox[65 165 550 730]{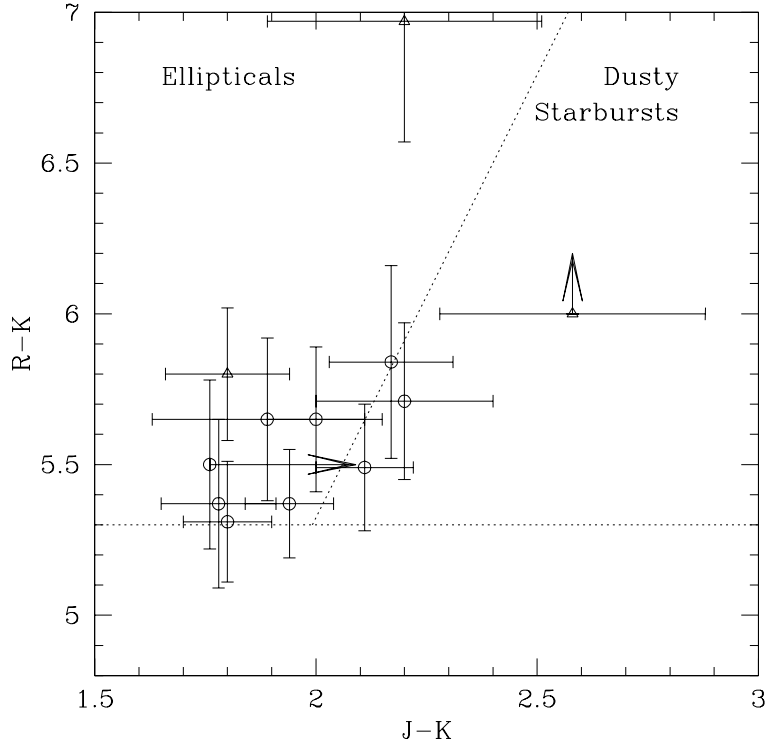}}
\caption{$J-K$ vs. $R-K$ diagram for EROs. The nine EROs are represented by
open circles and three EROs from the
literature by open triangles. The model by \citet{pozzetti00} to discriminate
between ellipticals and dusty starbursts ({\it dotted line})
nearly equally divides the ERO sample.
} \label{fig:rjk}
\end{figure*}

Colors are one means of breaking the degeneracy between ellipticals and
starbursts. The SED of an elliptical galaxy at $z \sim 1$ drops off sharply at
the 4000 \AA\ break between the $R$ and $K$ bands, while the SED of a dusty
starburst declines more gradually due to reddening.  As discussed by
\citet{pozzetti00}, observations between the $R$ and $K$ bands, such as $I$,
$z$, $J$, or $H$ can be used to measure the sharpness of the spectral break
and thus discriminate between these two scenarios. By this argument,
ellipticals appear bluer in $J-K$ than dusty starbursts.
Figure~\ref{fig:rjk}
shows $R-K$ vs.\ $J-K$ for the galaxies in Table~\ref{tbl:phot} ({\it open
circles}), where ellipticals lie to the left, towards bluer $J-K$ colors, and
dusty starbursts to the right, towards redder $J-K$ colors. \citet{pozzetti00}
convolved a range of models and observed galaxy spectral energy distributions
with $R$, $I$, $J$, and $H$ filters and found that ellipticals and dusty
starbursts separate in the $I-K$ vs.\ $J-K$ plane and $R-K$ vs.\ $J-K$ plane.
Their relation between $R-K$ and $J-K$ is shown in Figure~\ref{fig:rjk},
along with their suggested $R-K \geq 5.3$ selection criterion ({\it dotted
lines}). This color space does successfully classify three EROs from the 
literature with spectroscopic classifications, visible, and NIR photometry
({\it open triangles}): HR10 \citep{graham96}, LBDS 53W091 \citep{spinrad97}, 
and CL0939+4713B \citep{soifer99}. It is possible, however, that galaxies with 
a mix of old stars and dusty star formation could fall into either 
class on this diagram. For example, \citet{hall01} found evidence of this in 
their study of EROs associated with radio-loud quasars, where objects best 
fit by star formation and dust still had relatively blue $J-K$ colors. 
Of the nine DMS EROs shown in the Figure, seven have colors consistent with 
elliptical galaxies and two are consistent with dusty starbursts. However, 
six of the EROs fall within $1\sigma$ of the dividing line between these two 
classes and therefore either interpretation is consistent with the photometry. 

\subsection{SED Template Fits} \label{sec:sed}

Photometric redshifts with SED template fits are one way to expand on
simpler color discrimination techniques and solve for both the object redshift
and spectrophotometric galaxy type.  Photometric redshifts that fit SED
templates to photometric data rely on three pieces of information: a template
SED for the source object, the relative transmission of the system (filter,
detector, and optics) in each band, and a measurement of the source brightness
in that band.  Of these three quantities, the system transmission profile
should be the best-known quantity, in principle, and the true galaxy
SED the most uncertain. The templates used in photometric redshift codes
are either empirical SEDs, such as the set compiled by \citet{coleman80}, or
spectral synthesis templates, such as those based on the Bruzual \& Charlot
evolutionary code \citep[e.g. GISSEL98,][]{bruzual93}.

Photometric redshift codes are most accurate in the $z \sim 1$ -- 2
range when NIR photometry is available \citep[e.g.,][]{gwyn95,bolzonella00}.
At this range of redshifts, the strongest spectral feature, the 4000 \AA\
break, has shifted into the NIR region and the Lyman continuum
has not yet shifted into the ground-based $U$ filter bandpass.
Using mock galaxy catalogs produced with {\it hyperz} \citep{bolzonella00}, 
I measured the accuracy of the photometric redshift measurements for a 
$K\leq18$ mag galaxy sample with the same magnitude limits and noise properties 
as this survey in all nine filters. 
For a uniform distribution in redshifts, the scatter is $\sigma_z = 0.08$. 
The scatter for galaxies with 
$0.8 \leq z \leq 2$ is $\sigma_z = 0.1$, compared to $\sigma_z = 0.3$ for the 
same sample if only $UBVRI_{75}I_{86}$ photometry were available.  
The accuracy of photometric redshifts, and the associated best-fit templates,
depends on photometric quality as well as the number of filters. While the
EROs are all expected to be at $z \geq 0.8$ due to their $R-K$ colors,
they are often undetected in several filters (none are detected in $U$ or $B$)
and the photometric uncertainties in the remaining filters are generally
$> 0.1$ mag.  The larger the photometric errors, the less well contrained the
best-fit redshift, and particularly the best-fit galaxy template.

The empirical galaxy templates in {\it hyperz} are from \citet{coleman80} and 
represent the local galaxy population. 
The GISSEL98 galaxy templates are the 1998 update of the spectral synthesis 
models described by \citet{bruzual93}. 
All of the models have solar metallicity and a 
\citet{miller79} initial mass function.
Each model corresponds to a different present-day spectrophotometric galaxy
type. The elliptical template has all star formation occurring at high
redshift and thus represents classic passive luminosity evolution. The burst
model represents the opposite extreme with significant recent star formation.
The remaining galaxy types are represented by exponentially decaying star 
formation with different $e-$folding timescales. 

I used {\it hyperz} to find the best-fit empirical galaxy template, 
GISSEL98 elliptical template, and GISSEL98 starburst templates for each of 
the EROs listed in Table~\ref{tbl:phot}. The basic operation of this code 
is to take a series of input galaxy templates, vary the redshift and amount 
of reddening, and solve for the best combination of these quantities that 
match an input catalog of photometric measurements or upper limits. For these
model fits I adopted the \citet{calzetti00} reddening law. 
While there are a large number of filters predefined in {\it hyperz}, 
the DMS $I_{75}$ and $I_{86}$ filters are not included. Because of the 
importance of the system transmission profiles in each band, I added the 
profiles of the DMS filters from \citet{hall96a}, which include the detector 
response, and scans of the $J$, $H$, and $K$ filters from TIFKAM, which I 
convolved by a measurement of the atmospheric transmission at Kitt Peak. 

\begin{figure*}[t]
\centerline{
\epsfxsize=4truein
\epsfysize=4truein
\epsfbox[0 100 600 740]{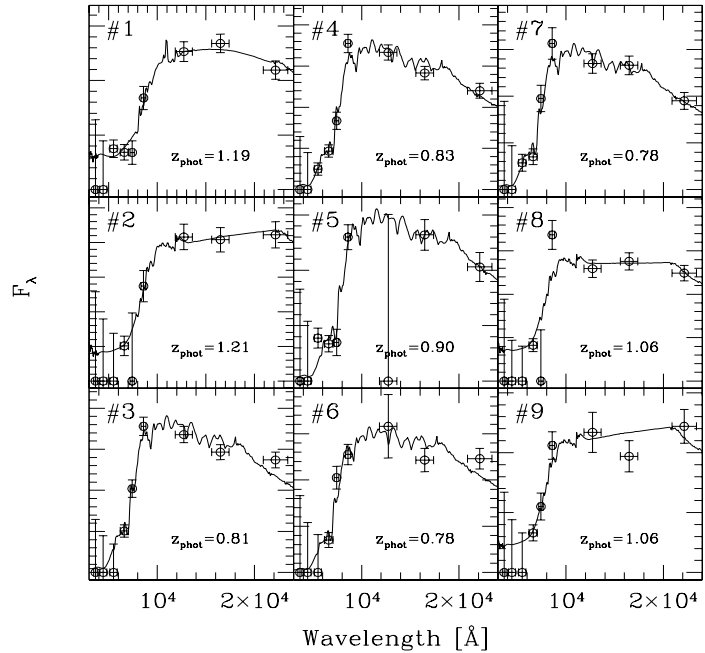}}
\caption{Best-fitting SEDs for the EROs. Each panel contains the
observed photometry for an ERO from Table~\ref{tbl:phot} ({\it solid
points with errorbars}) and the best-fit empirical galaxy template
({\it solid line}) from \citet{coleman80}.
} \label{fig:seds}
\end{figure*}

In all cases the photometric redshift predictions of the three fits 
shown in Table~\ref{tbl:zphot} are similar. 
One feature of the SED fits to these EROs is that on average
the best-fit elliptical template has a lower photometric redshift than the
best starburst template. This is because the dominant break in the SED of 
elliptical
galaxies is at 4000 \AA\, while the closest spectral
break in starburst galaxies is the Balmer continuum at 3650 \AA.
The rms scatter in the photometric redshift prediction for these
three fits is $\sigma_z < 0.1$. The photometric redshifts are thus quite 
robust and show clearly that these EROs are $z \geq 0.8$ galaxies. 
The relative quality of the three SED fits is comparable, with the difference
between the fits $\Delta \chi^{2}_{\nu} \leq 1$. These galaxies can therefore 
not be reliably classified as either ellipticals or dusty starbursts due to 
the large photometric errors. Most of the best fits with the \citet{coleman80} 
empirical SEDs are the elliptical template with no dust or spirals with some
dust. These fits suggest that most EROs are ellipticals, although old 
stellar populations at $z \sim 1$ are on order half the age of old stellar 
populations at $z\sim0$ and thus this empirical SED may not be an accurate 
representation of $z\sim1$ ellipticals. The GISSEL98 models do take the 
change in the age of the universe with redshift into account. The 
GISSEL98 elliptical templates fit the data with only a small amount of 
extinction, while the starburst models require on average $A_V = 1$ mag. 

\section{Discussion} \label{sec:dis}

The reported surface density of EROs has a large dispersion. \citet{elston88}
found two galaxies with $R-K\geq5$ and $K\leq18$ mag in 10 arcmin$^2$ for a 
surface density of 0.2~arcmin$^{-2}$. Figure~1 of \citet{thompson99} shows 
they found seven of these objects for a surface density of 0.05~arcmin$^{-2}$.
\citet{daddi00} surveyed 700~arcmin$^{2}$ and derived a surface density of 
$0.08$ arcmin$^{-2}$ for $R-K\geq5$ and $K\leq18$ mag. Their study also showed 
that EROs are clustered, which helps to explain the dispersion in
measurements of the surface density. In this survey I have identified nine 
EROs with $R-K\geq5.3$ and $K\leq18$ mag, which corresponds to a surface 
density of 0.049$^{+0.022}_{-0.016}$ arcmin$^{-2}$, where these uncertainties 
only correspond to the $1\sigma$ confidence limits \citep{gehrels86}. 
This surface density is nearly a factor of two higher than the surface density 
of 0.027 arcmin$^{-2}$ found by \citet{daddi00} for $R-K\geq5.3$ and 
$K\leq18$ mag (see Table~\ref{tbl:sigma}), although based on Poisson 
statistics these measurements are marginally consistent at the $1\sigma$ level. 

In addition to the random errors due to counting statistics, Eddington bias
and light lost outside the $R=2''$ aperture could systematically lower and 
raise the measured surface density, respectively. Eddington bias is important 
near the detection limit when objects with a steep number--magnitude relation 
are preferentially scattered into the sample and artificially enhance the 
observed space density. Based on the surface density vs. magnitude 
measurements published by \citet{daddi00}, the slope of the ERO 
number--magnitude relation (for $R-K\geq5.3$) is $\alpha \sim 0.9$ at 
$K \sim 18$ mag. The 
correction for the Eddington bias is $\sim 2.65 \sigma^2 \alpha^2$ and for 
a mean 
photometric uncertainty of $\sigma \sim 0.1$, the surface density may be 
overestimated by approximately 2\%. If the slope were a factor of two steeper, 
the overestimate is still less than 10\%. A more significant potential 
source of error arises if the EROs are large galaxies and a $R=2''$ aperture 
misses some fraction of their light. Then the surface density has been 
underestimated as effectively the photometric scale assigns them too faint 
a magnitude, thus the surface density at $K=18$ mag is really the surface 
density at $K=18-\delta m$ mag. The increase in surface density scales as 
$10^{\alpha\,\delta m}$, where for example $\delta m = 0.2$ ($r_h=1''$ for an 
exponential or $r^{1/4}$ profile) corresponds to an increase of 50\%, although 
a significant contribution from such large EROs appear to be ruled out by 
observations \citep{moriondo00}. 

While clustering will not change the true mean surface density, it will
increase the possible variation from survey to survey over the pure Poisson 
uncertainties quoted above. 
The DMS EROs are clearly strongly clustered as, for example, five of these
objects are in one of the subfields in Field01W (01WC, 01WC150W, and CF3), 
only 20\% of the total area with NIR data.  
The rms number counts of EROs in the DMS are $\sigma = 0.85$, compared to 
$\sigma_{\rm Poisson}=0.67$ for a surface density of $0.05$~arcmin$^2$ 
and an average area per field of 9~arcmin$^2$.
If I adopt the correlation amplitude $A = 0.024$ from \citet{daddi00} for 
$R-K\geq5$, $K\leq18$~mag (they had insufficient statistics to measure this 
quantity at $R-K\geq5.3$) the predicted rms counts \citep{roche99,daddi00} 
are $\sigma = 0.75$. The clustering signal in this suvey may be exaggerated 
by variations in the field to field sensitivity. 

The high surface density of EROs in this and other surveys provides a 
constraint for passive evolution and hierarchical galaxy formation models. 
Hierarchical models predict relatively few large, bright galaxies at 
$z\sim1$ as most these galaxies assemble at $z<1$. 
In a simple test of hierarchical galaxy formation, \citet{kauffmann98} 
predicted that less that 1\% of all galaxies with $16 < K < 18$ mag are at 
$z \geq 0.8$ and $\sim 10$\% of galaxies with $18 < K < 19$ mag are at 
$z \geq 0.8$ (see their Figure~4). Measurements of the $K$ number--magnitude 
relation \citep[e.g.][]{minezaki98b,martini01} show the galaxy surface density 
is $\sim 1.7$ arcmin$^{-2}$ for galaxies with $16 < K < 18$ mag and $\sim 3.3$ 
arcmin$^{-2}$ for galaxies with $18 < K < 19$ mag. As all EROs with 
spectroscopic or photometric redshifts are at $z \geq 0.8$, the surface 
density of EROs can be taken as a lower limit to the surface density of all 
$z \geq 0.8$ galaxies to test the 
\citet{kauffmann98} prediction of hierarchical galaxy formation. Their 
predicted fraction of galaxies at $z \geq 0.8$ implies that the surface 
density of EROs should be $< 0.017$ arcmin$^{-2}$ for $16 < K < 18$ mag, in 
conflict with the higher value I measure and the value measured by 
\citet{daddi00}. According the passive evolution model in 
\citet{kauffmann98}, $\sim$ 50\% of all galaxies at $K\leq18$ mag are at 
$z\geq0.8$, which corresponds to over an order of magnitude greater surface 
density of all $z\geq0.8$ galaxies than the ERO surface density presented 
here, even if the ERO surface density has been underestimated by as much as 
50\% due to lost light in the aperture. 

The ERO surface density I measure and also the measurement of 
\citet{daddi00} is still consistent with the results of 
$K$-selected spectroscopic and photometric redshift surveys 
\citep{songaila94,cowie96,fontana99}, which have
found a somewhat higher surface density of $z \geq 0.8$ galaxies than the
hierarchical model predictions. The cosmic variance
between different redshift and ERO surveys should be large given the 
clustering of EROs. The redshift distribution from a spectroscopic
survey is also most likely to be incomplete for the reddest and
highest-redshift galaxies and could underestimate the $z \sim 1$ contribution.

EROs provide such a good means to test the hierarchical galaxy formation 
model because their surface density is greater than the predicted surface 
density of all $z\geq0.8$ galaxies and they are only a lower limit on the total 
$z\geq0.8$ galaxy population in a $K-$selected galaxy survey. 
From the discussion of the systematic uncertainties in the surface density 
above, the overestimate due to Eddington bias is less significant than 
an underestimate due to lost galaxy light and therefore 0.05 arcmin$^{-2}$ 
may be a lower limit to the true ERO surface density. 
The surface density is therefore nearly a factor of three higher than 
the predicted surface density of all $z\geq0.8$ galaxies. 
Two other effects could bring the ERO surface density into better agreement 
with the hierarchical prediction, either by decreasing the observed number of 
EROs or increasing their expected number. Some fraction of EROs appear to be 
dusty starburst galaxies, rather than old ellipticals. While this 
fraction is very likely less than 50\% based on the colors of these EROs and 
spectroscopic observations of ERO samples from the literature 
\citep{cimatti99,cohen99}, decreasing the surface density from this survey 
by a factor of two still yields a higher surface 
density of objects than the hierarchical model. A more important effect, 
however, is that the current 
model predictions from \citet{kauffmann98} are for an $\om = 1$ universe. 
In a flat, $\Lambda-$dominated universe, which appears to be the best fit 
cosmological model, the comoving volume element vs.\ redshift is larger than 
in a flat, matter-dominated universe. For example, at $z=1$ a given 
surface area on the sky corresponds to a factor of 2.8 ($h_M$/$h_\Lambda$)$^3$ 
more comoving volume 
in an $\om = 0.3, \ol = 0.7$ universe than in an $\om = 1, \ol = 0$ universe. 
The volume per unit area on the sky is greater and implies a larger surface 
density of high-redshift objects than the matter-dominated model considered 
by \citet{kauffmann98}. 
The growth factor is also larger at fixed redshift in a $\Lambda-$dominated 
universe than in an $\om = 1$ model, which implies more large structures have 
formed and consequently a larger space density and surface density of bright 
galaxies.  The increase in surface density due to the change from a 
matter-dominated to a $\Lambda-$dominated universe is partially offset, 
however, by the increase in the luminosity distance. 

In this paper I have presented a new measurement of the ERO surface density 
and used photometric redshifts to show these objects are galaxies at 
$z\geq0.8$. The surface density of EROs is a lower limit to the total 
$z\geq0.8$ surface density, yet it is still a factor of three larger than the 
hierarchical galaxy formation model prediction for a matter-dominated universe. 
Hierarchical galaxy formation in a $\Lambda-$dominated universe, rather 
than a matter-dominated universe, could account for this discrepancy. 

\acknowledgements

I would like to thank Darren DePoy, Patrick Osmer, and David Weinberg for
helpful discussions and comments on this manuscript. In addition, helpful 
comments from an anonymous referee have improved and clarified this 
presentation.  I also acknowledge the staff of the MDM Observatory.  
I was supported in part by a Presidential Fellowship from 
Ohio State University and received additional travel and other support from a 
PEGS grant and the Department of Astronomy at Ohio State University.
TIFKAM was funded by the Ohio State University, the MDM consortium, MIT, and
NSF grant AST-9605012. NOAO and USNO paid for the development of the ALADDIN
arrays and contributed the array currently in use in TIFKAM.

{}

\clearpage

\begin{center}
\begin{deluxetable}{lcccccccccc}
\tabletypesize{\footnotesize}
\rotate
\tablenum{1}
\tablewidth{0pc}
\tablecaption{ERO Photometry \label{tbl:phot}}
\tablehead {
  \colhead{ERO} &
  \colhead{Field} &
  \colhead{$U$\tablenotemark{a}} &
  \colhead{$B$\tablenotemark{a}} &
  \colhead{$V$} &
  \colhead{$R$} &
  \colhead{$I_{75}$} &
  \colhead{$I_{86}$} &
  \colhead{$J$} &
  \colhead{$H$} &
  \colhead{$K$} \\
}
\startdata
1 & 01WC     & 23.0 & 23.8 & 23.440 (0.233) & 22.935 (0.266) & 22.904 (0.343) & 21.603 (0.138) & 19.377 (0.077) & 18.325 (0.069) & 17.376 (0.083) \\
2 & 01WC150W & 23.0 & 23.8 &  23.5\tablenotemark{a} & 23.258 (0.301) & 22.5\tablenotemark{a} & 21.826 (0.178) & 19.590 (0.096) & 18.622 (0.093) & 17.417 (0.098) \\
3 & 01WC150W & 23.0 & 23.8 & 23.5\tablenotemark{a} & 22.506 (0.159) & 21.710 (0.113) & 20.790 (0.069) & 19.072 (0.064) & 18.228 (0.068) & 17.132 (0.075) \\
4 & 01WC150W & 23.0 & 23.8 & 23.884 (0.323) & 22.591 (0.177) & 21.929 (0.136) & 20.794 (0.068) & 19.079 (0.061) & 18.266 (0.062) & 17.279 (0.082) \\
5 & 10EC     & 22.7 & 23.6 & 23.404 (0.251) & 22.949 (0.241) & 22.877 (0.372) & 21.133 (0.094) & 19.2\tablenotemark{a} & 18.343 (0.114) & 17.445 (0.139) \\
6 & 10EC     & 22.7 & 23.6 & 23.3\tablenotemark{a} & 22.906 (0.247) & 21.704 (0.130) & 21.153 (0.094) & 19.139 (0.236) & 18.433 (0.113) & 17.252 (0.098) \\
7 & 14NC150E & 22.6 & 23.8 & 23.732 (0.342) & 22.902 (0.260) & 21.758 (0.157) & 20.929 (0.177) & 19.307 (0.085) & 18.335 (0.081) & 17.532 (0.104) \\
8 & 14NC150E & 22.6 & 23.8 & 23.4\tablenotemark{a} & 22.727 (0.197) & 22.3\tablenotemark{a} &  20.849 (0.108) & 19.355 (0.084) & 18.293 (0.078) & 17.236 (0.078) \\
9 & CF3 & 23.0\tablenotemark{a} & 23.8\tablenotemark{a} & 23.5\tablenotemark{a} & 22.971 (0.223) & 22.385 (0.223) & 21.352 (0.117) & 19.464 (0.155) & 18.674 (0.145) & 17.261 (0.126) \\
\enddata
\tablenotetext{a}{$3\sigma$ upper limit.}
\tablecomments{Photometry of the ERO sample.
}
\end{deluxetable}
\end{center}

\begin{center}
\begin{deluxetable}{clcccccccccccc}
\tablecolumns{14}
\tabletypesize{\footnotesize}
\tablenum{2}
\tablewidth{0pt}
\tablecaption{ERO Photometric Redshifts\label{tbl:zphot}}
\tablehead{
\colhead{} & \colhead{} & \multicolumn{4}{c}{CWW} & \colhead{} &
\multicolumn{3}{c}{Elliptical} & \colhead{} & \multicolumn{3}{c}{Starburst} \\ 
\cline{3-6} \cline{8-10} \cline{12-14} \\
  \colhead{ERO} &
  \colhead{Field} &
  \colhead{$z$} &
  \colhead{$\chi^2_\nu$} &
  \colhead{$A_V$} &
  \colhead{SED} &
  \colhead{} &
  \colhead{$z$} & 
  \colhead{$\chi^2_\nu$} & 
  \colhead{$A_V$} & 
  \colhead{} &
  \colhead{$z$} &
  \colhead{$\chi^2_\nu$} &
  \colhead{$A_V$} 
}
\startdata
1 & 01WC     & 1.19 & 0.6 & 1.2 & Scd & & 1.17 & 1.0 & 0.4 & & 1.28 & 1.2 & 1.2 \\
2 & 01WC150W & 1.22 & 0.2 & 0.6 & Sbc & & 1.07 & 0.3 & 0.4 & & 1.06 & 0.2 & 1.0 \\
3 & 01WC150W & 0.81 & 1.3 & 0.0 & E   & & 0.81 & 1.3 & 0.2 & & 0.95 & 0.3 & 1.2 \\ 
4 & 01WC150W & 0.83 & 0.8 & 0.0 & E   & & 0.85 & 1.2 & 0.2 & & 0.96 & 0.4 & 1.0 \\
5 & 10EC     & 0.90 & 1.4 & 0.0 & E   & & 0.92 & 1.5 & 0.2 & & 1.17 & 0.9 & 1.2 \\
6 & 10EC     & 0.78 & 1.3 & 0.2 & E   & & 0.80 & 1.4 & 0.4 & & 0.91 & 0.8 & 1.2 \\
7 & 14NC150E & 0.78 & 0.7 & 0.0 & E   & & 0.78 & 0.8 & 0.2 & & 0.79 & 0.7 & 0.0 \\
8 & 14NC150E & 1.07 & 1.7 & 0.4 & Sbc & & 0.84 & 1.6 & 0.4 & & 0.98 & 0.9 & 1.2 \\
9 & CF3      & 1.07 & 0.7 & 0.6 & Sbc & & 0.90 & 0.9 & 0.4 & & 1.00 & 0.4 & 1.2 \\
\enddata
\tablecomments{Photometric results for the ERO sample. Columns 1 \& 2 contain 
the number and field for each ERO as in Table~\ref{tbl:phot}. Columns 3 
-- 6 list the best-fit photometric redshift, $\chi^2_\nu$, $A_V$, and 
SED for the \citet{coleman80} SED templates, columns 7 -- 9 the best-fit 
parameters for a GISSEL98 elliptical, and columns 10 -- 12 the best-fit 
parameters for a GISSEL98 starburst model.
}
\end{deluxetable}
\end{center}

\begin{center}
\begin{deluxetable}{lccccc}
\tablenum{3}
\tablewidth{0pt}
\tablecaption{ERO Surface Density\tablenotemark{a}\, ($R-K>5.3$)\label{tbl:sigma}}
\tablehead{
  \colhead{} &
  \colhead{$\Sigma_K$ (all galaxies)} &
  \colhead{$z\geq0.8$ fraction} &
  \colhead{$\Sigma_K$ ($z\geq0.8$)} &
  \colhead{$\Sigma_{ERO}$\tablenotemark{b}} &
  \colhead{$\Sigma_{ERO}$\tablenotemark{c}} \\
}
\startdata
$16 < K < 18$ mag 	& 1.67 & $<$ 1\%  & $< 0.017$ & 0.05 & 0.027 \\
$18 < K < 19$ mag 	& 3.33 & $<$ 10\% & $< 0.33$  &      & 0.27  \\
\enddata
\tablenotetext{a}{arcmin$^2$}
\tablenotetext{b}{this paper}
\tablenotetext{c}{\citet{daddi00}}
\tablecomments{
Surface density of EROs with $R-K\geq5.3$. Column~1 lists the two $K$ 
magnitude ranges and column~2 the surface density of all galaxies in this 
magnitude range from \citet{martini01}. The percent of all galaxies at 
$z\geq0.8$ in these magnitude ranges from the hierarchical galaxy model 
prediction of \citet{kauffmann98} is given in column~3. 
Column~4 lists the expected surface density of galaxies at $z\geq0.8$, while 
columns~5 \& 6 list the measured surface density of EROs, which are lower 
limits to the surface density of all $z\geq0.8$ galaxies. 
The measurements in these columns are higher than the hierarchical model 
predictions for $K\leq18$ mag, but are consistent with this model for 
$18 < K < 19$ mag. 
}
\end{deluxetable}
\end{center}

\end{document}